\documentclass[aps,prd,twocolumn,showpacs,superscriptaddress]{revtex4-1}
\usepackage{amsmath}
\usepackage[dvips]{graphicx}

\begin{document}

\title{An amplitude analysis of the \boldmath$\pi^{0}\pi^{0}$ system produced in radiative \boldmath$J/\psi$ decays}

\date{June 1, 2015}

\author{
\begin{small}
\begin{center}
M.~Ablikim$^{1}$, M.~N.~Achasov$^{9,f}$, X.~C.~Ai$^{1}$, O.~Albayrak$^{5}$, M.~Albrecht$^{4}$, D.~J.~Ambrose$^{44}$, A.~Amoroso$^{48A,48C}$, F.~F.~An$^{1}$, Q.~An$^{45,a}$, J.~Z.~Bai$^{1}$
% ...
\\
\vspace{0.2cm}
(BESIII Collaboration)\\
\vspace{0.2cm}
A.~P.~Szczepaniak$^{19,53,54}$, P.~Guo$^{19,53}$\\
\vspace{0.2cm} {\it
$^{1}$ Institute of High Energy Physics, Beijing 100049, People's Republic of China\\
% ...
$^{4}$ Bochum Ruhr-University, D-44780 Bochum, Germany\\
$^{5}$ Carnegie Mellon University, Pittsburgh, Pennsylvania 15213, USA\\
% ...
$^{9}$ G.I. Budker Institute of Nuclear Physics SB RAS (BINP), Novosibirsk 630090, Russia\\
% ...
$^{19}$ Indiana University, Bloomington, Indiana 47405, USA\\
% ...
$^{44}$ University of Rochester, Rochester, New York 14627, USA\\
$^{45}$ University of Science and Technology of China, Hefei 230026, People's Republic of China\\
% ...
$^{48}$ (A)University of Turin, I-10125, Turin, Italy; (B)University of Eastern Piedmont, I-15121, Alessandria, Italy; (C)INFN, I-10125, Turin, Italy\\
% ...
$^{53}$ Center for Exploration of Energy and Matter, Indiana University, Bloomington, IN 47403, USA\\
$^{54}$ Theory Center, Thomas Jefferson National Accelerator Facility, Newport News, VA 23606, USA\\
\vspace{0.2cm}
$^{a}$ Also at State Key Laboratory of Particle Detection and Electronics, Beijing 100049, Hefei 230026, People's Republic of China\\
% ...
$^{f}$ Also at the Novosibirsk State University, Novosibirsk, 630090, Russia\\
% ...
}\end{center}
\vspace{0.4cm}
\end{small}
}

\begin{abstract}
An amplitude analysis of the $\pi^{0}\pi^{0}$ system produced in radiative $J/\psi$ decays is 
presented. In particular, a piecewise function that describes the dynamics of the $\pi^{0}\pi^{0}$ 
system is determined as a function of $M_{\pi^{0}\pi^{0}}$ from an analysis of the 
$(1.311\pm0.011)\times10^{9}$ $J/\psi$ decays collected by the BESIII detector.  The goal of this 
analysis is to provide a description of the scalar and tensor components of the $\pi^0\pi^0$ system 
while making minimal assumptions about the properties or number of poles in the amplitude.  Such a 
model-independent description allows one to integrate these results with other related results from 
complementary reactions in the development of phenomenological models, which can then be used 
to directly fit experimental data to obtain parameters of interest.  The branching fraction of 
$J/\psi \to \gamma \pi^{0}\pi^{0}$ is determined to be $(1.15\pm0.05)\times10^{-3}$, where the 
uncertainty is systematic only and the statistical uncertainty is negligible.
\end{abstract}


\maketitle

\section{Introduction} \label{Introduction}

While the Standard Model of particle physics has yielded remarkable successes, the connection 
between the quantum chromodynamics (QCD) and the complex structure of hadron dynamics 
remains elusive. The light isoscalar scalar meson spectrum ($I^{G}J^{PC}=0^{+}0^{++}$), for 
example, remains relatively poorly understood despite many years of investigation. This lack of 
understanding is due in part to the presence of broad, overlapping states, which are poorly described 
by the most accessible analytical methods (see the ``Note on scalar mesons below 2~GeV'' in the 
PDG)~\cite{PDBook}. The PDG reports eight $0^{+}0^{++}$ mesons, which have widths between 100 
and 450~MeV. Several of these states, including the $f_{0}(1370)$, are characterized in the PDG 
only by ranges of values for their masses and widths.

Knowledge of the low mass scalar meson spectrum is important for several reasons. In particular, the 
lightest glueball state is expected to have scalar quantum numbers~\cite{B93,MP99,C06,O13}. The 
existence of such a state is an excellent test of QCD. Experimental observation of a glueball state 
would provide evidence that gluon self-interactions can generate a massive meson. Unfortunately, 
glueballs may mix with conventional quark bound states, making the identification of glueball states 
experimentally challenging. The low mass scalar meson spectrum is also of interest in probing the 
fundamental interactions of hadrons in that it allows for testing of Chiral Perturbation Theory to one 
loop~\cite{PY05}.

The scalar meson spectrum has been studied in many reactions, including $\pi$$N$ 
scattering~\cite{G01}, $p\bar{p}$ annihilation~\cite{A96}, central hadronic production~\cite{B99}, 
decays of the $\psi'$~\cite{A07}, $J/\psi$~\cite{A05,A04,A042}, $B$~\cite{L12}, $D$~\cite{B07}, and 
$K$~\cite{B072} mesons, $\gamma\gamma$ formation~\cite{M07} and $\phi$ radiative 
decays~\cite{A02}. In particular, a coupled channel analysis using the K-matrix formalism has been 
performed using data from pion production, $p\bar{p}$ and $n\bar{p}$ annihilation, and $\pi\pi$ 
scattering~\cite{AS03}. Similar investigations would benefit from the inclusion of data from radiative 
$J/\psi$ decays, which provide a complementary source of hadronic production.

An attractive feature of a study of the two pseudoscalar spectrum in radiative $J/\psi$ decays is the 
relative simplicity of the amplitude analysis. Conservation of parity in strong and electromagnetic 
interactions, along with the conservation of angular momentum, restricts the quantum numbers of the 
pseudoscalar-pseudoscalar pair. Only amplitudes with even angular momentum and positive parity 
and charge conjugation quantum numbers are accessible ($J^{PC}=0^{++}, 2^{++}, 4^{++},$ etc). 
Initial studies suggest that only the $0^{++}$ and $2^{++}$ amplitudes are significant in radiative 
$J/\psi$ decays to $\pi^{0}\pi^{0}$. The neutral channel ($\pi^{0}\pi^{0}$) is of particular interest due 
to the lack of sizable backgrounds like $\rho\pi$, which present a challenge for an analysis of the 
charged channel ($\pi^{+}\pi^{-}$)~\cite{A06}.

Radiative $J/\psi$ decays to $\pi^{+}\pi^{-}$ have been analyzed previously by the MarkIII~\cite{B87}, 
DM2~\cite{A87}, and BES~\cite{B96} experiments. Decays to $\pi^{0}\pi^{0}$ were also studied at 
Crystal Ball~\cite{KW89} and BES~\cite{B98}, but these analyses were severely limited by statistics, 
particularly for the higher mass states. Each of these analyses reported evidence for the 
$f_{2}(1270)$ and some possible additional states near 1.710~GeV/$c^{2}$ and 2.050~GeV/$c^{2}$. 
More recently, the BESII experiment studied these channels and implemented a partial wave 
analysis~\cite{A06}. Prominent features in the results include the $f_{2}(1270)$, $f_{0}(1500)$, 
and $f_{0}(1710)$. However, this analysis, like its predecessors, was limited by complications from 
large backgrounds and low statistics. Due to statistical limitations, the $\pi^{0}\pi^{0}$ channel was 
used only as a cross check on the analysis of the charged channel.

Historically, amplitude analyses like that in Ref.~\cite{A06} have relied on modeling the
$s$-dependence of the $\pi\pi$ interaction, where $s$ is the invariant mass squared of the two pions, 
as a coherent sum of resonances, each described by a Breit-Wigner function. In doing so, a model is 
built whose parameters are resonance properties, e.g. masses, widths and branching fractions. A 
correspondence exists between these properties and the residues and poles of the $\pi\pi$ scattering 
amplitude in the complex $s$ plane; however, this correspondence is only valid in the limit of an 
isolated narrow resonance that is far from open thresholds ({\it cf.} Ref.~\cite{PDBook}).  For regions 
containing multiple overlapping resonances with large widths and the presence of thresholds, all of 
which occur in the $0^{++}$ $\pi\pi$ spectrum, an amplitude constructed from a sum of Breit-Wigner 
functions becomes an approximation.  While such an approximation provides a practical and 
controlled way to parameterize the data -- additional resonances can be added to the sum until an 
adequate fit is achieved -- it is unknown how well it maintains the correspondence between Breit-
Wigner parameters and the analytic structure of the $\pi\pi$ amplitude that one seeks to study, 
{\it i.e.,} the fundamental strong interaction physics.  Often statistical precision, a lack of 
complementary constraining data, or a limited availability of models leaves the simple Briet-Wigner 
sum as a necessary but untested assumption in analyses, thereby rendering the numerical result 
only useful in the context of that assumption. In the context of this paper we refer to the Breit-Wigner 
sum as a ``mass dependent fit", that is, the model used to fit the data has an assumed $s$ 
dependence.

In this analysis we exploit the statistical precision provided by $(1.311\pm0.011)\times10^{9}$ 
$J/\psi$ decays collected with the BESIII detector~\cite{YZ14,NJpsi} to measure the components of 
the $\pi\pi$ amplitude independently for many small regions of $\pi\pi$ invariant mass, which allows 
one to construct a piecewise complex function from the measurements that describes the $s$-
dependence of the $\pi\pi$ dynamics.  Such a construction makes minimal assumptions about the 
$s$-dependence of the $\pi\pi$ interaction.  We refer to this approach in the context of the paper as a 
``mass independent fit". 

The mass independent approach has some drawbacks.  First, due to the large number of bins, one is 
left with a set of about a thousand parameters that describe the amplitudes with no single parameter 
tied to an individual resonance of interest.  Second, mathematical ambiguities result in multiple sets 
of optimal parameters in each mass region.  If only $J=0$ and $J=2$ resonances are significant, 
there are two ambiguous solutions.  However, in general, if one includes $J\ge 4$ the number of 
ambiguous solutions increases resulting in multiple allowed piecewise functions.  Finally, in order to 
make the results practically manageable for subsequent analysis, the assumption of Gaussian errors 
must be made -- an assumption that cannot be validated in general.  Similar limitations are present in 
other analyses of this type, {\it e.g.,} Ref.~\cite{G01}.  In spite of these limitations, which are discussed 
further in Appendices~\ref{sec:amb} and~\ref{supplement} the results of the mass independent 
amplitude analysis presented here represent a measurement of $\pi\pi$ dynamics in radiative 
$J/\psi$ decays that minimizes experimental artifacts and potential systematic biases due to 
theoretical assumptions. The results are presented with the intent of motivating the development 
of dynamical models with reaction independent parameters that can subsequently be optimized 
using experimental data.  All pertinent information for the use of these results in the study of 
pseudoscalar-pseudoscalar dynamics is included in the supplemental materials 
(Appendix~\ref{supplement}).

\section{The BESIII Detector} \label{Detector}

The Beijing Spectrometer (BESIII) is a general-purpose, hermetic detector located at the Beijing 
Electron-Positron Collider (BEPCII) in Beijing, China. BESIII and BEPCII represent major upgrades to 
the BESII detector and BEPC accelerator. The physics goals of the BESIII experiment cover a broad 
research program including charmonium physics, charm physics, light hadron spectroscopy and 
$\tau$ physics, as well as searches for physics beyond the standard model. The detector is described 
in detail elsewhere~\cite{BESDet}. A brief description follows.

The BESIII detector consists of five primary components working in conjunction to facilitate the 
reconstruction of events. A superconducting solenoid magnet provides a uniform magnetic field 
within the detector. The field strength was 1.0~T during data collection in 2009, but was reduced to 
0.9~T during the 2012 running period. Charged particle tracking is performed with a helium-gas 
based multilayer drift chamber (MDC). The momentum resolution of the MDC is expected to be better 
than 0.5\% at 1~GeV/c, while the expected dE/dx resolution is 6\%. With a timing resolution of 80~ps 
(110~ps) in the barrel (endcap), a plastic scintillator time-of-flight (TOF) detector is useful for particle 
identification. The energies of electromagnetic showers are determined using information from the 
electromagnetic calorimeter (EMC).  The EMC consists of $6240$ CsI(Tl) crystals arranged in one 
barrel and two endcap sections. With an angular coverage of about 93\% of 4$\pi$, the EMC provides 
an energy resolution of 2.5\% (5\%) at 1.0~GeV and a position resolution of 6~mm (9~mm) in the 
barrel (endcap). Finally, particles that escape these detectors travel through a muon chamber system 
(MUC), which provides additional information on the identity of particles. The MUC provides 2~cm 
position resolution for muons and covers 89\% of 4$\pi$. Muons with momenta over 0.5~GeV are 
detected with an efficiency greater than 90\%. The efficiency of pions reaching the MUC is about 
10\% at this energy.

Selection criteria and background estimations are studied using a \textsc{geant4} Monte Carlo (MC) 
simulation. The BESIII Object Oriented Simulation Tool (\textsc{boost})~\cite{boost} provides a 
description of the geometry, material composition, and detector response of the BESIII detector. The 
MC generator \textsc{kkmc}~\cite{kkmc} is used for the production of $J/\psi$ mesons by $e^{+}e^{-}$ 
annihilation, while \textsc{besevtgen}~\cite{evtgen} is used to generate the known decays of the 
$J/\psi$ according to the world average values from the PDG~\cite{PDBook}. The unknown portion of 
the $J/\psi$ decay spectrum is generated with the Lundcharm model~\cite{lundcharm}.

\section{Event Selection} \label{Event}

In order to be included in the amplitude analysis, an event must have at least five photon candidates 
and no charged track candidates. Any photon detected in the barrel (endcap) portion of the EMC 
must have an energy of at least 25 (50)~MeV. Four of the five photons are grouped into two pairs that 
may each originate from a $\pi^{0}$ decay. The invariant mass of any photon pair associated with a 
$\pi^{0}$ must fall within 13~MeV/$c^{2}$ of the $\pi^{0}$ mass. A 6C kinematic fit is performed on 
each permutation of photons to the final state $\gamma\pi^{0}\pi^{0}$. This includes a constraint on 
the four-momentum of the final state to that of the initial $J/\psi$ (4C) and an additional constraint 
(1C) on each photon pair to have an invariant mass equal to that of a $\pi^{0}$.

Significant backgrounds in this channel include $J/\psi$ decays to $\gamma\eta$ 
($\eta\rightarrow\pi^{0}\pi^{0}\pi^{0}$) and $\gamma\eta'$ 
($\eta'\rightarrow\eta\pi^{0}\pi^{0}; \eta\rightarrow\gamma\gamma$). Restricting the $\chi^{2}$ from 
the 6C kinematic fit is an effective means of reducing the backgrounds of this type. Events with a 
$\pi^{0}\pi^{0}$ invariant mass, $M_{\pi^{0}\pi^{0}}$, below KK threshold (the region in which 
these backgrounds are significant) must have a $\chi^{2}$ less than 20. Events above KK threshold 
need only have a $\chi^{2}$ less than 60. To reduce the background from $J/\psi$ 
decays to $\omega\pi^{0}$ ($\omega\rightarrow\gamma\pi^{0}$), the invariant mass of each 
$\gamma\pi^{0}$ pair is required to be at least 50~MeV/$c^{2}$ away from the $\omega$ 
mass~\cite{PDBook}. Finally, in order to reduce the misreconstructed background arising from pairing 
the radiated photon with another photon in the event to form a $\pi^{0}$, the invariant mass of the 
radiated photon paired with any $\pi^{0}$ daughter photon is required to be greater than 0.15 
GeV/c$^{2}$.

If more than one permutation of five photons in an event satisfy these selection criteria, only the 
permutation with the minimum $\chi^{2}$ from the 6C kinematic fit is retained. After all event selection 
criteria are applied, the number of events remaining in the data sample is 442,562. MC studies 
indicate that the remaining backgrounds exist at a level of about 1.8\% of the size of the total sample. 
Table~\ref{table:bkgs} lists the major backgrounds. 

\begin{figure*}[htp]
\includegraphics[width=\textwidth]{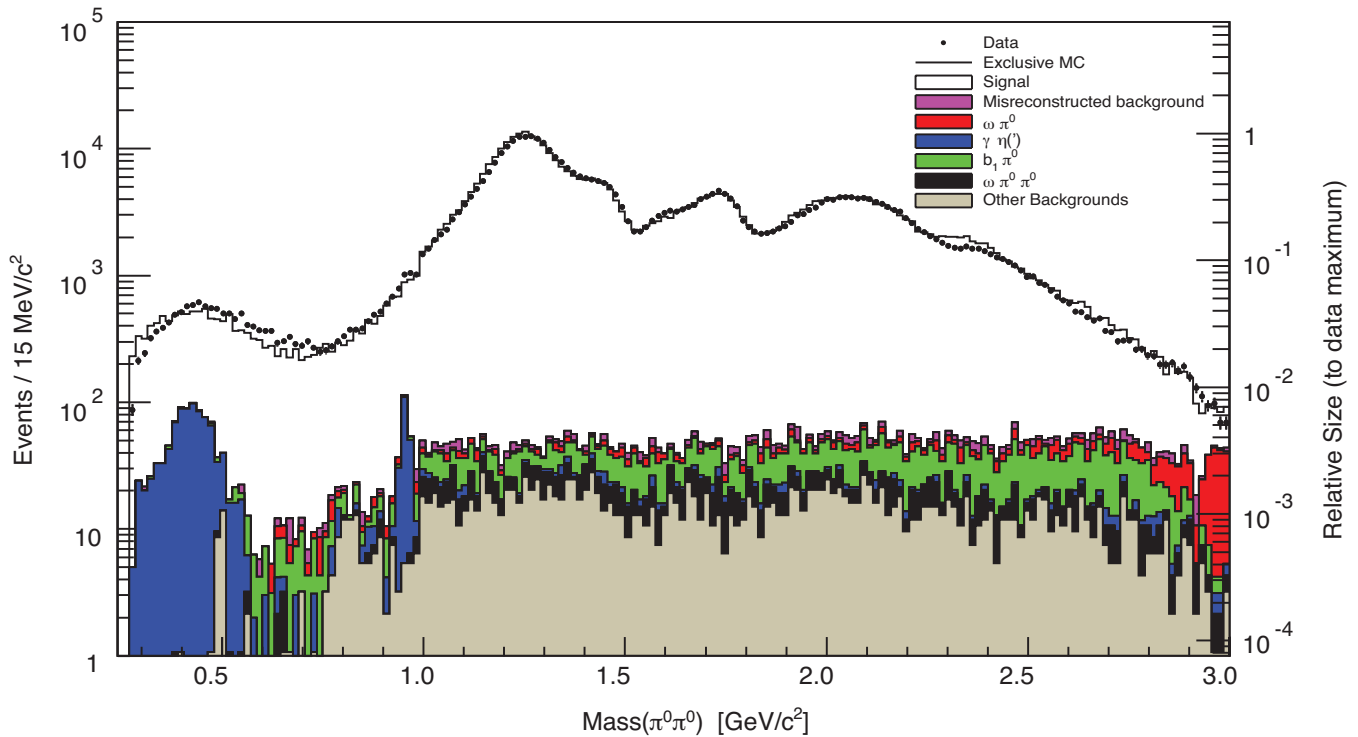}
\caption{\label{fig:pi0pi0im} The $M_{\pi^{0}\pi^{0}}$ spectrum after all selection criteria have been 
applied. The black markers represent the data, while the histograms depict the backgrounds 
according to the MC samples. The signal (white) and misreconstructed background (pink) are 
determined from an exclusive MC sample that resembles the data. The other backgrounds are 
determined from an inclusive MC sample (see Table~\ref{table:bkgs}). The components of the 
stacked histogram from bottom up are unspecified backgrounds, $\omega\pi^{0}\pi^{0}$, 
$b_{1}\pi^{0}$, $\gamma\eta(')$, $\omega\pi^{0}$, the misreconstructed background, and the signal.}
\end{figure*}

Backgrounds from $J/\psi$ decays to $\gamma\eta(')$ are well understood and are studied with an 
exclusive MC sample, which is generated according to the PDG branching fractions for these 
reactions. Other backgrounds are studied using an inclusive MC sample generated using 
\textsc{besevtgen}, with the exception of the misreconstructed background, which is studied using an 
exclusive MC sample that resembles the data. The latter MC sample was generated using a set of 
Breit-Wigner resonances with couplings determined from a mass dependent fit to the data sample. 
The $M_{\pi^{0}\pi^{0}}$ spectrum after all selection criteria have been applied is shown in 
Fig.~\ref{fig:pi0pi0im}. The reconstruction efficiency is determined to be 28.7\%, according to the 
results of the mass independent amplitude analysis. Continuum backgrounds are investigated with a 
data sample collected at a center of mass energy of 3.080~GeV. The continuum backgrounds are 
scaled by luminosity and a correction factor for the difference in cross section as a function of center 
of mass energy. When scaled by luminosity, only 3,632 events, which represents approximately 
0.8\% of the signal, survive after all signal isolation requirements.

\begin{table}[ht]
\centering
\caption[The number of events remaining after each signal isolation criterion]{\label{table:bkgs} The 
number of events remaining after all selection criteria for each of a number of background reactions 
is shown in the right column. The backgrounds are broken into three groups. The first group contains 
the signal mimicking decays. The second lists the remaining backgrounds from $J/\psi$ decays to 
$\gamma\eta(')$, while the third group lists a few additional backgrounds. The backgrounds explicitly 
listed here represent about 93\% of the total background according to the MC samples. The 
misreconstructed background includes those events in which one of the daughter photons from a 
$\pi^{0}$ decay is taken as the radiated photon.}
\begin{tabular}{c c}
\hline\hline
Decay channel & Number of events  \\
\hline\hline
$J/\psi \rightarrow \gamma\pi^{0}\pi^{0}$ (data) & 442,562 \\
$e^{+}e^{-} \rightarrow \gamma\pi^{0}\pi^{0}$ (continuum) & 3,632 \\
\hline
$J/\psi \rightarrow b_{1}\pi^{0}; b_{1} \rightarrow \gamma\pi^{0}$ & 1,606 \\
$J/\psi \rightarrow \omega\pi^{0}; \omega \rightarrow \gamma\pi^{0}$ & 865 \\
$J/\psi \rightarrow \rho\pi^{0}; \rho \rightarrow \gamma\pi^{0}$ & 778 \\
Misreconstructed background & 608 \\
\\
$J/\psi \rightarrow \gamma\eta; \eta \rightarrow 3\pi^{0}$ & 903 \\
$J/\psi \rightarrow \gamma\eta'; \eta' \rightarrow \eta\pi^{0}\pi^{0}; \eta \rightarrow \gamma\gamma$ & 377 \\
\\
$J/\psi \rightarrow \omega\pi^{0}\pi^{0}; \omega \rightarrow \gamma\pi^{0}$ & 775 \\
$J/\psi \rightarrow b_{1}\pi^{0}; b_{1} \rightarrow \omega\pi^{0}; \omega \rightarrow \gamma\pi^{0}$ & 578 \\
$J/\psi \rightarrow \omega\eta; \omega \rightarrow \gamma\pi^{0}$ & 409 \\
$J/\psi \rightarrow \omega f_{2}(1270); \omega \rightarrow \gamma\pi^{0}$ & 299 \\
$J/\psi \rightarrow \gamma\eta_{c}; \eta_{c} \rightarrow \gamma\pi^{0}\pi^{0} or \pi^{0}\pi^{0}\pi^{0}$ & 255 \\
\\
Other backgrounds & 507 \\
\hline
Total Background (MC) & 7,960 \\
\hline\hline
\end{tabular}
\end{table}

\section{Amplitude Analysis} \label{Analysis}

\subsection{General Formalism}

The results of the mass independent amplitude analysis of the $\pi^{0}\pi^{0}$ system are obtained 
from a series of unbinned extended maximum likelihood fits. The amplitudes for radiative $J/\psi$ 
decays to $\pi^{0}\pi^{0}$ are constructed in the radiative multipole basis, as described in detail in 
Appendix~\ref{sec:amps}. 

Let $U^{M,\lambda_{\gamma}}$ represent the amplitude for radiative $J/\psi$ decays to 
$\pi^{0}\pi^{0}$,
\begin{equation}
  U^{M,\lambda_{\gamma}}(\vec{x},s) = \langle \gamma\pi^{0}\pi^{0}|H|J/\psi \rangle
\end{equation}
where $\vec{x}=\{\theta_{\gamma},\phi_{\gamma},\theta_{\pi},\phi_{\pi}\}$ is the position in phase 
space, $s = M_{\pi^{0}\pi^{0}}^{2}$ is the invariant mass squared of the $\pi^{0}\pi^{0}$ pair, $M$ is 
the polarization of the $J/\psi$, and $\lambda_{\gamma}$ is the helicity of the radiated photon. For 
the reaction under study the possible values of both $M$ and $\lambda_{\gamma}$ are $\pm1$. The 
amplitude may be factorized into a piece that contains the radiative transition of the $J/\psi$ to an 
intermediate state $X$ and a piece that contains the QCD dynamics
\begin{equation} \label{eq:factorized}
\begin{split}
  U^{M,\lambda_{\gamma}}(\vec{x},s) = \sum_{j,J_{\gamma},X}&\langle \pi^{0}\pi^{0}|H_{QCD}|X_{j,J_{\gamma}} \rangle \\
  &\times \langle \gamma X_{j,J_{\gamma}}|H_{EM}|J/\psi \rangle,
\end{split}
\end{equation}
where $j$ is the angular momentum of the intermediate state and $J_{\gamma}$ indexes the 
radiative multipole transitions. The sum over $X$ includes any pseudoscalar-pseudoscalar final 
states ($\pi\pi$, $K\bar{K}$, etc) that may rescatter into $\pi^{0}\pi^{0}$. We assume that the 
contribution of the 4$\pi$ final state to this sum is negligible, with the result that rescattering effects 
become important only above the $K\bar{K}$ threshold.

The amplitude in Eq.~\eqref{eq:factorized} may be further factorized by pulling out the angular 
distributions,
\begin{equation} \label{eq:couplings}
\begin{split}
  U^{M,\lambda_{\gamma}}(\vec{x},s)=\sum_{j,J_{\gamma},X}&T_{j,X}(s)\Theta_{j}^{M,\lambda_{\gamma}}(\theta_{\pi},\phi_{\pi}) \\
&\times g_{j,J_{\gamma},X}(s)\Phi_{j,J_{\gamma}}^{M,\lambda_{\gamma}}(\theta_{\gamma},\phi_{\gamma}),
\end{split}
\end{equation}
where $g_{j,J_{\gamma},X}(s)$ is the coupling for the radiative decay to intermediate state $X$. The 
functions $\Theta_{j}^{M,\lambda_{\gamma}}(\theta_{\pi},\phi_{\pi})$ and 
$\Phi_{j,J_{\gamma}}^{M,\lambda_{\gamma}}(\theta_{\gamma},\phi_{\gamma})$ contain the angular 
dependence of the decay of the $X$ to $\pi^{0}\pi^{0}$ and the radiative $J/\psi$ decay, respectively. 
The part of the amplitude that describes the $\pi^{0}\pi^{0}$ dynamics is the complex function 
$T_{j,X}(s)$, which is of greatest interest for this study. However, this function cannot be separated 
from the coupling $g_{j,J_{\gamma},X}(s)$. Instead the product is measured according to
\begin{equation} \label{eq:factorization}
  V_{j,J_{\gamma}}(s) \approx \sum_{X} g_{j,J_{\gamma},X}(s) T_{j,X}(s).
\end{equation}
This product will be called the coupling to the state with characteristics $j,J_{\gamma}$. Note here 
that, if rescattering effects are assumed to be minimal (the only possible $X$ is $\pi\pi$), all 
amplitudes with the same $j$ have the same phase. The effect of rescattering is to break the 
factorizability of Eq.~\eqref{eq:factorization}. Finally, the amplitude may be written
\begin{equation}
  U^{M,\lambda_{\gamma}}(\vec{x},s)=\sum_{j,J_{\gamma}}V_{j,J_{\gamma}}(s)A_{j,J_{\gamma}}^{M,\lambda_{\gamma}}(\vec{x}),
\end{equation}
where $A_{j,J_{\gamma}}^{M,\lambda_{\gamma}}(\vec{x})$ contains the piece of the amplitude that 
describes the angular distributions and is determined by the kinematics of an event.

Any amplitude with total angular momentum greater than zero will have three components (the 
$0^{++}$ amplitude has only an E1 component). Thus, three $2^{++}$ amplitudes, relating to E1, M2, 
and E3 radiative transitions, are included in the analysis. While any amplitude with even total angular 
momentum and positive parity and charge conjugation is accessible for this decay, studies show that 
the $4^{++}$ amplitude is not significant in this region. In particular, no set of four continuous 
15 MeV/c$^{2}$ bins yield a difference in $-2 \ln{L}$ greater than 28.8 units, which corresponds to a 
five sigma difference, under the inclusion of a $4^{++}$ amplitude. As no narrow spin-4 states are 
known, this suggests that only the $0^{++}$ and $2^{++}$ amplitudes are significant. The systematic 
uncertainty due to ignoring a $4^{++}$ amplitude that may exist in the data is described below in 
Sec.~\ref{sec:4pp}.

\subsection{Parameterization}

The dynamical function in Eq.~\eqref{eq:factorization} may be parameterized in various ways. A 
common parameterization, discussed in the introduction, is a sum of interfering Breit-Wigner 
functions,
\begin{equation} \label{eq:amplitude}
  V_{j,J_{\gamma}}(s) = \sum_{\beta} k_{j,J_{\gamma},\beta}BW_{j,J_{\gamma},\beta}(s),
\end{equation}
where $BW_{j,J_{\gamma},\beta}(s)$ represents a Breit-Wigner function with characteristics (mass 
and width) $\beta$ and strength $k_{j,J_{\gamma},\beta}$.  

To avoid making such a strong model dependent assumption, we choose to bin the data sample as a 
function of $M_{\pi^{0}\pi^{0}}$ and to assume that the part of the amplitude that describes the 
dynamical function is constant over a small range of $s$,
\begin{equation} \label{eq:miparam}
  U^{M,\lambda_{\gamma}}(\vec{x},s)=\sum_{j,J_{\gamma}}V_{j,J_{\gamma}} A_{j,J_{\gamma}}^{M,\lambda_{\gamma}}(\vec{x}).
\end{equation}

For the scenario posed in Eq.~\eqref{eq:miparam}, the couplings may be taken as the free 
parameters of an extended maximum likelihood fit in each bin of $M_{\pi^{0}\pi^{0}}$. It is then 
possible to extract a table of complex numbers (the free parameters in each bin) that describe the 
dynamical function of the $\pi^{0}\pi^{0}$ interaction.

The intensity function, $I(\vec{x})$, which represents the density of events at some position in phase 
space $\vec{x}$, is given by
\begin{equation} \label{eq:intensity}
  I(\vec{x})=\sum_{M,\lambda_{\gamma}}\left|\sum_{j,J_{\gamma}}V_{j,J_{\gamma}} A_{j,J_{\gamma}}^{M,\lambda_{\gamma}}(\vec{x})\right|^{2}.
\end{equation}
The incoherent sum includes the observables of the reaction (which are not measured). For the 
reaction under study, the observables are the polarization of the $J/\psi$, $M=\pm1$, and the helicity 
of the radiated photon, $\lambda_{\gamma}=\pm1$. The free parameters are constrained to be the 
same in each of the four pieces of the incoherent sum.

In the figures and supplemental results that follow, the intensity of the amplitude in each bin is 
reported as a number of events corrected for acceptance and detector efficiency.  That is, for the bin 
of $M_{\pi^{0}\pi^{0}}$ indexed by $k$ and bounded by $s_k$ and $s_{k+1}$ (the boundaries in $s$ 
of the bin) we report, for each amplitude indexed by $j$ and $J_\gamma$, the quantity
\begin{equation}
I_{j,J_\gamma}^{k} = \int_{s_k}^{s_{k+1}}\sum_{M,\lambda_{\gamma}}\left|V^k_{j,J_{\gamma}} A_{j,J_{\gamma}}^{M,\lambda_{\gamma}}(\vec{x})\right|^2~d\vec{x}.
\end{equation}
In practice, we absorb the size of phase space into the fit parameters.  In doing so we fit for 
parameters $\widetilde{V}^k_{j,J_\gamma}$ which are the $V^k_{j,J_\gamma}$ scaled by the square 
root of the size of phase space in bin $k$.

\subsection{Background subtraction}

The mass independent amplitude analysis treats each event in the data sample as a signal event. 
For a clean sample, the effect of remaining backgrounds should be small relative to the statistical 
errors on the amplitudes. However, the backgrounds from $J/\psi$ decays to $\gamma\eta(')$ 
introduce a challenge. Both of these backgrounds peak in the low mass region near interesting 
structures. The background from $J/\psi$ decays to $\gamma\eta$ lies in the region of the
$f_{0}(500)$, which is of particular interest for its importance to Chiral Perturbation 
Theory~\cite{PDBook,K03}. The $\gamma\eta'$ background peaks near the $f_{0}(980)$, which is 
also of particular interest due to its strong coupling to $K\bar{K}$ and its implications for a scalar 
meson nonet~\cite{AT04}. Therefore, the effect of these backgrounds is removed by using a 
background subtraction method.

If a data sample is entirely free of backgrounds, the likelihood function is constructed as
\begin{equation}
  L(\vec{\xi})=\prod_{i=1}^{N_{\mathrm{data}}^{\mathrm{sig}}}f(\vec{x}_{i}|\vec{\xi}),
\end{equation}
where $f(\vec{x}|\vec{\xi})$ is the probability density function (pdf) to observe an event with a 
particular set of kinematics $\vec{x}$ and parameters $\vec{\xi}=\{\widetilde{V}^k_{j,J_{\gamma}}\}$. 
The total number of parameters in the mass independent analysis is 1,178 (seven times the number 
of bins above $K\bar{K}$ threshold and five times the number of bins below $K\bar{K}$ threshold). 
The number of events in the pure data sample is given by $N_{\mathrm{data}}^{\mathrm{sig}}$.

Now, the likelihood may be written
\begin{equation} \label{compositepdf}
  L(\vec{\xi})=\prod_{i=1}^{N_{\mathrm{data}}^{\mathrm{sig}}}f(\vec{x}_{i}|\vec{\xi})\prod_{j=1}^{N_{\mathrm{data}}^{\mathrm{bkg}}}f(\vec{x}_{j}|\vec{\xi})\prod_{k=1}^{N_{\mathrm{data}}^{\mathrm{bkg}}}f(\vec{x}_{k}|\vec{\xi})^{-1},
\end{equation}
where an additional likelihood, which describes the reaction for background events, has been 
multiplied and divided. Consider now a more realistic data sample that consists not only of signal 
events, but also contains some number of background events, $N_{\mathrm{data}}^{\mathrm{bkg}}$. 
Then the product of the first two factors of Eq.~\eqref{compositepdf} are simply the likelihood for the 
entire (contaminated) data sample, but the overall likelihood represents only that of the pure signal 
since the background likelihood has been divided. For a given data set, any backgrounds remaining 
after selection criteria have been applied are difficult to distinguish from the true signal. Rather than 
using the true background to determine the background likelihood, it is therefore necessary to 
approximate it with an exclusive MC sample. That is,
\begin{equation}
  \prod_{i=1}^{N_{\mathrm{data}}^{\mathrm{bkg}}}f(\vec{x}_{i}|\vec{\xi})^{-1}\approx
  \prod_{i=1}^{N_{\mathrm{MC}}^{\mathrm{bkg}}}f(\vec{x}_{i}|\vec{\xi})^{-w_{i}},
\end{equation}
where the weight, $w_{i}$, is necessary for scaling purposes. For example, if the MC sample is twice 
the size of the expected background, a weight factor of 0.5 is necessary. Finally, the likelihood 
function may be written
\begin{equation} \label{eq:sublik}
  L(\vec{\xi})=\prod_{i=1}^{N_{\mathrm{data}}}f(\vec{x}_{i}|\vec{\xi})\prod_{j=1}^{N_{\mathrm{MC}}^{\mathrm{bkg}}}f(\vec{x}_{j}|\vec{\xi})^{-w_{j}}.
\end{equation}
In practice, this likelihood distribution is multiplied by a Poisson distribution for the extended 
maximum likelihood fits such that
\begin{equation} \label{eq:extendedlik}
  L(\vec{\xi})=\frac{e^{-\mu}\mu^{N_{\mathrm{data}}}}{N_{\mathrm{data}}!}
	\prod_{i=1}^{N_{\mathrm{data}}}f(\vec{x}_{i}|			\vec{\xi})\prod_{j=1}^{N_{\mathrm{MC}}^{\mathrm{bkg}}}f(\vec{x}_{j}|\vec{\xi})^{-w_{j}}.
\end{equation}

An exclusive MC sample for the backgrounds due to $J/\psi$ decays to $\gamma\eta(')$ is generated 
according to the branching fractions given by the PDG~\cite{PDBook}. This MC sample is required to 
pass all of the selection criteria that are applied to the data sample. Any events that remain are 
included in the unbinned extended maximum likelihood fit with a negative weight ($-w_{j}=-1$ in 
Eq.~\eqref{eq:sublik}). In this way, the inclusion of the MC sample in the fit approximately cancels the 
effect of any remaining backgrounds of the same type in the data sample.

\subsection{Ambiguities} \label{sec:ambiguities}

Another challenge to the amplitude analysis is the presence of ambiguities. Since the intensity 
function, which is fit to the data, is constructed from a sum of absolute squares, it is possible to identify 
multiple sets of amplitudes which give identical values for the total intensity. In this way, multiple 
solutions may give comparable values of $-2 \ln{L}$ for a particular fit. For this particular analysis, two 
types of ambiguities are present. Trivial ambiguities arise due to the possibility of the overall 
amplitude in each bin to be rotated by $\pi$ or to be reflected over the real axis in the complex plane. 
These may be partially addressed by applying a phase convention to the results of the fits. Non-trivial 
ambiguities arise from the freedom of amplitudes with the same quantum numbers to have different 
phases. The non-trivial ambiguities represent a greater challenge to the analysis and cannot be 
eliminated without introducing model dependencies.

While it is not possible in principle to measure the absolute phase of the amplitudes, it is possible to 
study the relative phases of individual amplitudes. Therefore in each of the fits, one of the amplitudes 
(the $2^{++}$ E1 amplitude) is constrained to be real. The phase difference between the other 
amplitudes and that which is constrained can then be determined in each mass bin.

As mentioned above, a set of trivial ambiguities arises due to the possibility of the overall amplitude 
in each bin to be rotated by $\pi$ or to be reflected over the real axis in the complex plane. Each of 
these processes leave the intensity distribution unchanged. This issue is partially resolved by 
establishing a phase convention in which the amplitude that is constrained to be real is also 
constrained to be positive. The remaining ambiguity is related to the inability to determine the 
absolute phase. The phase of the total amplitude may change sign without inducing a change in the 
total intensity. Therefore, when a phase difference approaches zero, it is not possible to determine if 
the phase difference should change sign. The amplitude analysis results are presented here with the 
arbitrary convention that the phase difference between the $0^{++}$ amplitude and the $2^{++}$ E1 
amplitude is required to be positive. One may invert the sign of this phase difference in a given bin, 
but then all other phase differences in that bin must also be inverted.

The presence of non-trivial ambiguities is attributed to rescattering effects, which allow for amplitudes 
with the same quantum numbers, $J^{PC}$, to have different phases. The couplings, 
$g_{j,J_{\gamma},X}(s)$, in Eq.~\eqref{eq:factorization} are real functions of $s$. Since the dynamical 
amplitude, $T_{j,X}(s)$, does not depend on $J_{\gamma}$, its phase is the same for each of the 
amplitudes with the same $J^{PC}$ (in particular, the $2^{++}$ E1, M2 and E3 amplitudes). However, 
if more than one intermediate state, $X$, is present, differences between the couplings to these 
amplitudes may result in a phase difference. Therefore, in the region above the $K\bar{K}$ threshold 
the $2^{++}$ amplitudes may have different phases. However, below $K\bar{K}$ threshold the 
phases of these amplitudes are constrained to be the same. That is, rescattering through 4$\pi$ is 
assumed to be negligible.

By writing out the angular dependence of the intensity function, it is possible to show that the freedom 
to have phase differences between the components of a given amplitude ($2^{++}$ E1, M2, and E3, 
for example) generates an ambiguity in the intensity distribution. For this channel and considering 
only $0^{++}$ and $2^{++}$ amplitudes, two non-trivial ambiguous solutions may be present in each 
bin above $K\bar{K}$ threshold. The knowledge of one solution can be used to mathematically 
predict its ambiguous partner. In fact, some bins do not exhibit multiple solutions, but have a 
degenerate ambiguous pair. A study of these ambiguities (Appendix~\ref{sec:amb}) shows 
consistency between the mathematically predicted and experimentally determined ambiguities. Both 
ambiguous solutions are presented, because it is impossible to know which represent the physical 
solutions without making some additional model dependent assumptions. If more than two solutions 
are found in a given bin, all solutions within 1 unit of log likelihood from the best solution are 
compared to the predicted value {\it derived from the best solution} and only that which matches the 
prediction is accepted as the ambiguous partner.

\subsection{Results} \label{Results}

\subsubsection{Amplitude intensities and phases}

\begin{figure*}[htp]
\includegraphics[width=0.9\textwidth]{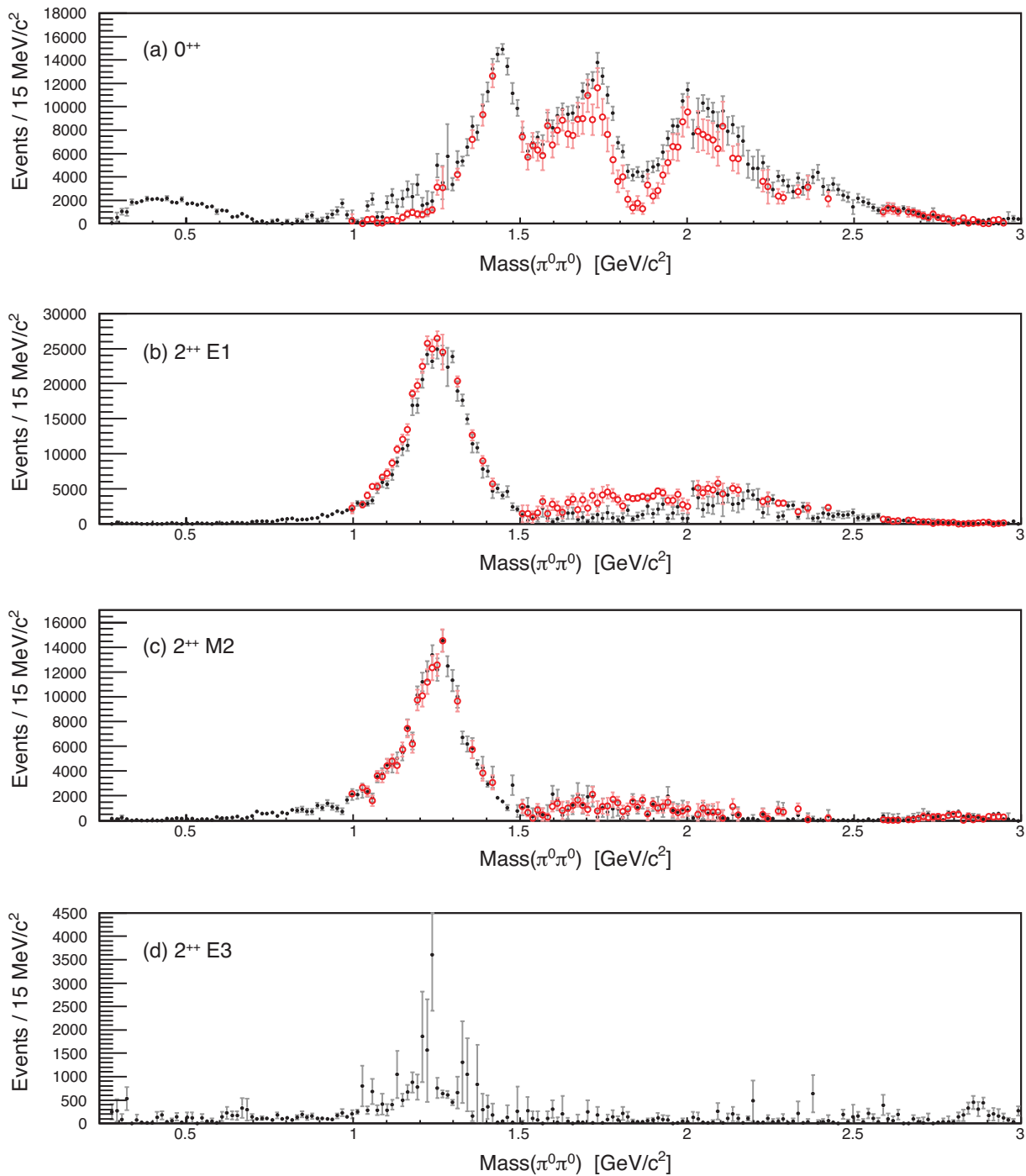}
\caption{\label{fig:jpsiinten_nom} The intensities for the (a) $0^{++}$, (b) $2^{++}$ E1, (c) $2^{++}$ 
M2 and (d) $2^{++}$ E3 amplitudes as a function of $M_{\pi^{0}\pi^{0}}$ for the nominal results. The 
solid black markers show the intensity calculated from one set of solutions, while the open red 
markers represent its ambiguous partner. Note that the intensity of the $2^{++}$ E3 amplitude is 
redundant for the two ambiguous solutions (see Appendix~\ref{sec:amb}). Only statistical errors are 
presented.}
\end{figure*}

\begin{figure*}[htp]
\includegraphics[width=1.0\textwidth]{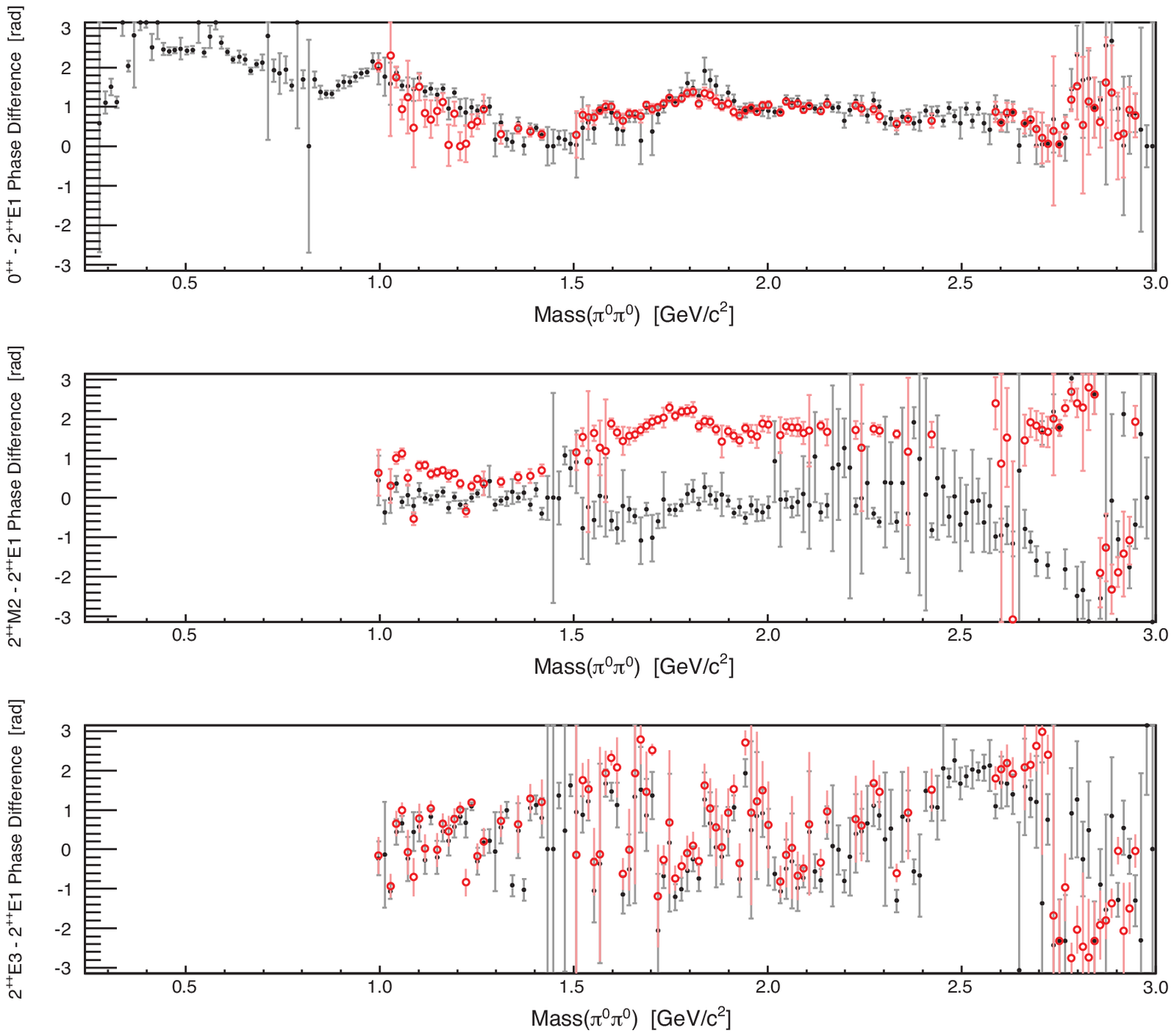}
\caption{\label{fig:jpsiinten_phases} The phase differences relative to the reference amplitude 
($2^{++}$ E1) for the (a) $0^{++}$, (b) $2^{++}$ M2, and (c) $2^{++}$ E3 amplitudes as a function of 
$M_{\pi^{0}\pi^{0}}$ for the nominal results. The solid black markers show the phase differences 
calculated from one set of solutions, while the open red markers represent the ambiguous partner 
solutions. An arbitrary phase convention is applied here in which the phase difference between the 
$0^{++}$ and $2^{++}$ E1 amplitudes is required to be positive. Only statistical errors are presented.}
\end{figure*}

The intensity for each amplitude as a function of $M_{\pi^{0}\pi^{0}}$ is plotted in 
Fig.~\ref{fig:jpsiinten_nom}. Each of the phase differences with respect to the reference amplitude 
($2^{++}$ E1), which is constrained to be real, is plotted in Fig.~\ref{fig:jpsiinten_phases}. Above the 
$K\bar{K}$ threshold, two distinct sets of solutions are apparent in most bins as expected. The bins 
below about 0.6~GeV/$c^{2}$ also contain multiple solutions, but with different likelihoods and are 
attributed to local minima in the likelihood function. The nominal solutions below 0.6~GeV/$c^{2}$ 
are determined by requiring continuity in each intensity and phase difference as a function of 
$M_{\pi^{0}\pi^{0}}$. Only statistical errors are presented in the figures.

It is apparent that the ambiguous sets of solutions in the nominal results are distinct in some regions, 
while they approach and possibly cross at other points. The most powerful discriminator of this effect 
is the phase difference between the E1 and M2 components of the $2^{++}$ amplitude (see the 
middle plot of Fig.~\ref{fig:jpsiinten_phases}). Regions in which the solutions may cross are apparent 
at 0.99~GeV/$c^{2}$, near 1.3~GeV/$c^{2}$, and above 2.3~GeV/$c^{2}$. Since the results in each 
bin are independent of their neighbor, it is not possible to identify two distinct, smooth solutions at 
these crossings. 

\subsubsection{Discussion}

The results of the mass independent analysis exhibit significant structures in the $0^{++}$ amplitude 
just below 1.5~GeV/$c^{2}$ and near 1.7~GeV/$c^{2}$.  This region is where one might expect to 
observe the the states $f_0(1370)$, $f_0(1500)$, and $f_0(1710)$ which are often cited as being 
mixtures of two scalar light quark states and a scalar glueball~\cite{CK01,C00}.  A definitive 
statement on the number and properties of the scattering amplitude poles in this region of the 
spectrum requires model-dependent fits to the data. The effectiveness of any such model-dependent 
study could be greatly enhanced by including similar data from the decay $J/\psi\to\gamma KK$ in an 
attempt to isolate production features from partial widths to $KK$ and $\pi\pi$ final states.

Additional structures are present in the $0^{++}$ amplitude below 0.6~GeV/$c^{2}$ and near 
2.0~GeV/$c^{2}$. It seems reasonable to interpret the former as the $\sigma$ ($f_{0}(500)$).  The 
latter could be attributed to the $f_{0}(2020)$. The presence of the four states below 2.1~GeV/$c^{2}$ 
would be consistent with the previous study of radiative $J/\psi$ decays to $\pi\pi$ by 
BESII~\cite{A06}. Finally, the results presented here also suggest two possible additional structures 
in the $0^{++}$ spectrum that were not observed in Ref.~\cite{A06}. These include a structure just 
below 1~GeV/$c^{2}$, which may indicate an $f_{0}(980)$, but the enhancement in this region is 
quite small. There also appears to be some structure in the $0^{++}$ spectrum around 
2.4~GeV/$c^{2}$.

In the $2^{++}$ amplitude, the results of this analysis indicate a dominant contribution from what 
appears to be the $f_{2}(1270)$, consistent with previous results~\cite{A06}. However, the remaining 
structure in the $2^{++}$ amplitude appears significantly different than that assumed in the model 
used to obtain the BESII results~\cite{A06}. In particular, the region between 1.5 and 
2.0~GeV/$c^{2}$ was described in the BESII analysis with a relatively narrow $f_{2}(1810)$. One 
permutation of the nominal results (the red markers in Fig.~\ref{fig:jpsiinten_nom}) indicates that the 
structures in this region are much broader, while the other permutation (the black markers in 
Fig.~\ref{fig:jpsiinten_nom}) suggests that there is very little contribution from any $2^{++}$ states in 
this region.

The tensor spectrum near 2~GeV/$c^{2}$ is of interest in the search for a tensor glueball.  Previous 
investigations of the $J/\psi\to\gamma\pi^0\pi^0$ channel reported evidence for a narrow 
($\Gamma\approx 20$~MeV) tensor glueball candidate, $f_J(2230)$~\cite{B98}.  While a 
model-dependent fit is required to place a limit on the production of such a state using these data, we 
note that based on the reported value of $B(J/\psi\to\gamma f_J(2230))$~\cite{B96}, one would 
naively expect to observe a peak for the $f_J(2230)$ with an integral that is of order $4\times10^{5}$ 
but concentrated only in roughly two bins of $M(\pi^0\pi^0)$, corresponding to the full width of the 
$f_J(2230)$. Such a structure seems difficult to accommodate in the extracted $2^{++}$ amplitude.

\subsection{Branching fraction} \label{BR}

The results of the mass independent amplitude analysis allow for a measurement of the branching 
fraction of radiative $J/\psi$ decays to $\pi^{0}\pi^{0}$, which is determined according to:
\begin{equation} \label{eq:br}
\mathcal{B}(J/\psi\rightarrow\gamma\pi^{0}\pi^{0}) = \frac{N_{\gamma\pi^{0}\pi^{0}}-N_{\mathrm{bkg}}}{\epsilon_{\gamma}N_{J/\psi}},
\end{equation}
where $N_{\gamma\pi^{0}\pi^{0}}$ is the number of acceptance corrected events, 
$N_{\mathrm{bkg}}$ is the number of remaining background events, $\epsilon_{\gamma}$ is an 
efficiency correction necessary to extrapolate the $\pi^{0}\pi^{0}$ spectrum down to a radiative 
photon energy of zero, and $N_{J/\psi}$ is the number of $J/\psi$ decays in the data. The number of 
acceptance corrected events is determined from the amplitude analysis by summing the total 
intensity from each $M_{\pi^{0}\pi^{0}}$ bin. The number of remaining background events is 
determined according to the inclusive and exclusive MC samples. The fractional background 
contamination in each bin $i$, $R_{\mathrm{bkg},i}$, is determined before acceptance correction. 
The number of background events is then determined by assuming $R_{\mathrm{bkg,i}}$ is constant 
after acceptance correction such that the number of background events in bin $i$, 
$N_{\mathrm{bkg},i}$, is given by the product of $R_{\mathrm{bkg},i}$ and the number of acceptance 
corrected events in the same bin, $N_{\gamma\pi^{0}\pi^{0},i}$. Note that the backgrounds from to 
$J/\psi$ decays to $\gamma\eta(')$ are removed during the fitting process and are not included in this 
factor. The efficiency correction factor, $\epsilon_{\gamma}$, is determined by calculating the fraction 
of phase space that is removed by applying the selection requirements on the energy of the radiative 
photon. This extrapolation increases the total number of events by 0.07\%. Therefore, 
$\epsilon_{\gamma}$ is taken to be 0.9993.

The backgrounds remaining after event selection fall into three categories. The misreconstructed 
backgrounds are determined from an exclusive MC sample that resembles the data. Events that 
remain in a continuum data sample taken at 3.080~GeV after selection criteria have been applied are 
also taken as a background. Finally, the other remaining backgrounds are determined using the 
inclusive MC sample. Each of these backgrounds is scaled appropriately. In total, the acceptance 
corrected number of background events, $N_{\mathrm{bkg}}$, is determined to be 35,951. The 
number of radiative $J/\psi$ decays to $\pi^{0}\pi^{0}$, $N_{\gamma\pi^{0}\pi^{0}}$, is determined to 
be 1,543,050 events. The branching fraction for this decay is then determined to be 
$(1.151\pm0.002)\times10^{-3}$, where the error is statistical only.

\section{Systematic Uncertainties}

The systematic uncertainties for the mass independent analysis include two types. First, the 
uncertainty due to the effect of backgrounds from $J/\psi$ decays to $\gamma\eta(')$ are addressed 
by repeating the analysis and treating the background in a different manner.  The second type of 
systematic uncertainty is that due to the overall normalization of the results. Sources of systematic 
uncertainties of this type include the photon detection efficiency, the total number of $J/\psi$ decays, 
the effect of various backgrounds, differences in the effect of the kinematic fit between the data and 
MC samples and the effect of model dependencies. The uncertainty on the branching fraction of 
$\pi^{0}$ to $\gamma\gamma$ according to the PDG is 0.03\%~\cite{PDBook}, which is negligible in 
relation to the other sources of uncertainty. The systematic uncertainties are described below and 
summarized in Table~\ref{tab:eff}. These uncertainties also apply to the branching fraction 
measurement. Finally, several cross checks are also performed.

\subsection{ \boldmath $J/\psi\to\gamma\eta$ and $J/\psi\to\gamma\eta^\prime$ Background Uncertainty} \label{sec:alt2}

The amplitude analysis is performed with the assumption that all backgrounds have been eliminated. 
Studies using Monte Carlo simulation indicate this is a valid assumption for most of the 
$M_{\pi^{0}\pi^{0}}$ spectrum.  However, significant backgrounds from $J/\psi$ decays to 
$\gamma\eta$ and $\gamma\eta'$ exist in many mass bins below about 1~GeV/$c^{2}$. Rather than 
inflating the errors of these bins according to the uncertainty introduced by these backgrounds, which 
would not take into account the bin-to-bin correlations, a set of alternate results is presented in which 
the $\gamma\eta(')$ backgrounds are not subtracted. 

The fraction of events in $J/\psi$ decays to $\gamma\eta(')$ that survive the event selection criteria for 
the $\gamma\pi^{0}\pi^{0}$ final state is very small (about 0.02\%). Minor changes to the modeling of 
these decays may therefore have a large effect on the backgrounds. The  difference between the 
nominal results and the alternate results, which treat the backgrounds differently, can be viewed as 
an estimator of the systematic error in the results due to these backgrounds.  

\begin{figure*}[htp]
\includegraphics[width=\textwidth]{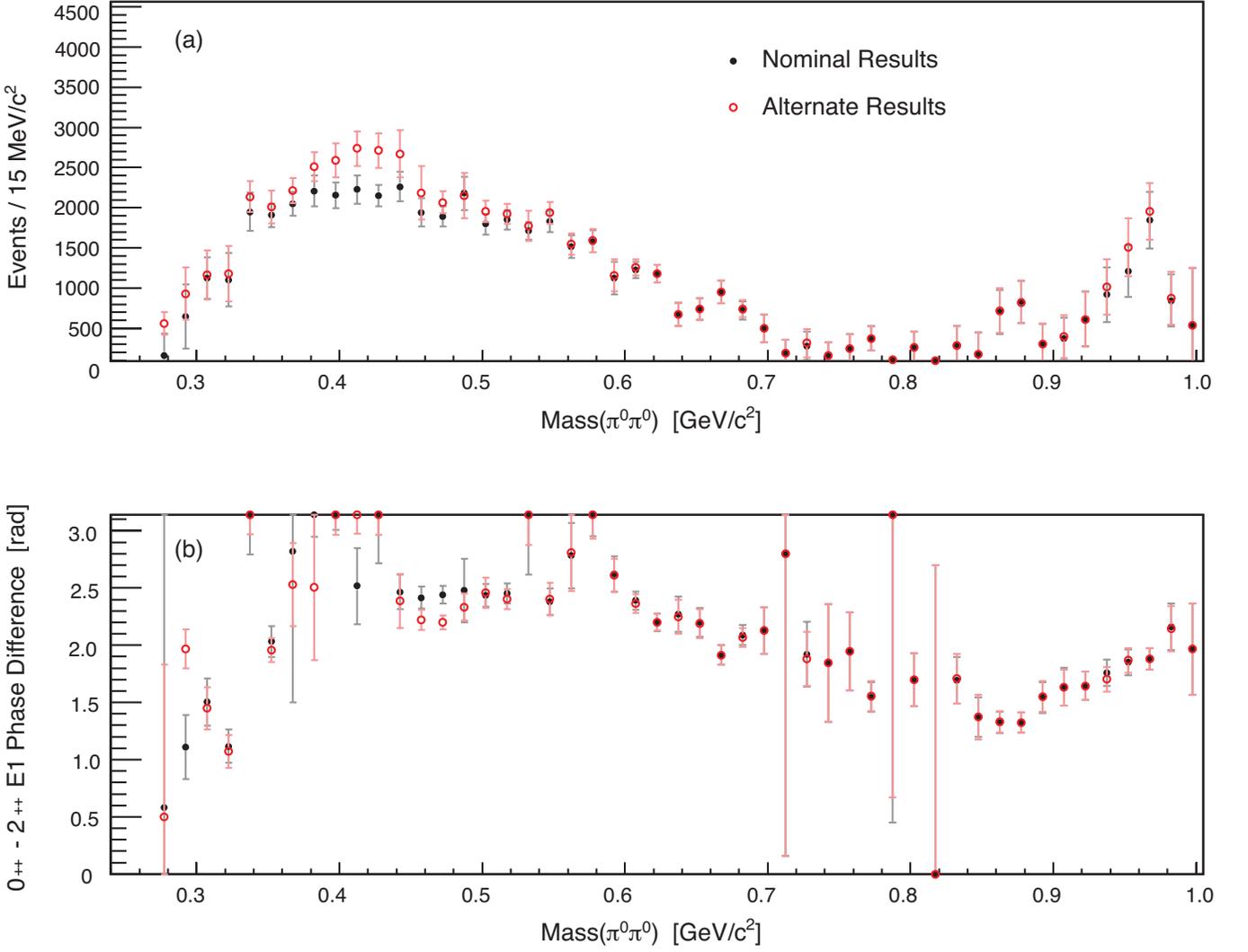}
\caption{\label{fig:alt2_all} A comparison of the (a) $0^{++}$ intensity and (b) phase difference 
relative to the $2^{++}$ E1 amplitude for the nominal results and the alternate results, in which the 
$\gamma\eta(')$ backgrounds have not been subtracted from the data. The solid black markers show 
the nominal results, while the red markers represent the alternate results. Only statistical errors are 
presented.}
\end{figure*}

The distinctive feature of the alternate results is an enhancement in the $0^{++}$ intensity in the 
region below about 0.6~GeV/$c^{2}$ and near the $\eta'$ peak. This may be interpreted as the 
contribution of the events from $J/\psi$ decays to $\gamma\eta(')$, which are being treated as signal 
events. A comparison of the $0^{++}$ amplitude for nominal results and the alternate results is 
presented in Fig.~\ref{fig:alt2_all}. The results for the other amplitudes are consistent between the two 
methods.  Any conclusion drawn from these data that is sensitive to choosing specifically the 
alternate or nominal results is not a robust conclusion.

\subsection{Uncertainties in the overall normalization}

\subsubsection{Photon Detection Efficiency}

The primary source of systematic uncertainty for this analysis comes from the reconstruction of 
photons. To account for this uncertainty, the photon detection efficiency of the BESIII detector is 
studied using the so called tag and probe method on a sample of $J/\psi$ decays to 
$\pi^{+}\pi^{-}\pi^{0}$, where the $\pi^{0}$ decays into two photons. One of these final state photons is 
reconstructed, along with the two charged tracks, while the other photon is left as a missing particle in 
the event. This information can then be used to determine the region in the detector where the 
missing photon is expected. The photon detection efficiency is calculated by taking the ratio of the 
number of missing photons that are detected in this region to the number that are expected. The 
numbers of detected and expected photons are determined with fits to the two photon invariant mass 
distributions.

The systematic uncertainty due to photon reconstruction is determined by investigating the 
differences between the photon detection efficiencies of the inclusive MC sample and that of the data 
sample. This difference is measured to be less than 1.0\%, which is taken to be the systematic 
uncertainty per photon. For the five photon final state the overall uncertainty due to this effect is 
therefore taken to be 5.0\%.

An additional source of uncertainty, which is due to mismodelling of the photon detection efficiency 
as a function of the angular and energy dependence of the radiative photon, was studied using the 
same channel. The phase space MC samples used for normalization in each bin of the mass 
independent amplitude analysis were modified to account for differences in the photon detection 
efficiency between the data and inclusive MC samples. The mass independent analysis was then 
repeated using the modified phase space MC samples. Neither the differences in angular nor energy 
dependence had a significant effect on the results of the analysis. The effects of mismodelling of this 
type are therefore taken to be negligible.

\subsubsection{Number of $J/\psi$}

The number of $J/\psi$ decays is determined from an analysis of inclusive hadronic events
\begin{equation} \label{eq:njpsi}
  N_{J/\psi}=\frac{N_{\mathrm{sel}}-N_{\mathrm{bg}}}{\epsilon_{\mathrm{trig}}\times\epsilon_{\mathrm{data}}^{\psi(2S)}\times f_{\mathrm{cor}}},
\end{equation}
where $N_{\mathrm{sel}}$ represents the number of inclusive events remaining after selection 
criteria have been applied and $N_{\mathrm{bg}}$ is the number of background events estimated 
with a data sample collected at 3.080~GeV. The efficiency for the trigger is given by 
$\epsilon_{\mathrm{trig}}$, while $\epsilon_{\mathrm{data}}^{\psi(2S)}$ is the detection efficiency for 
$J/\psi$ inclusive decays determined from $\psi(2S)$ decays to $\pi^{+}\pi^{-}J/\psi$. Finally, 
$f_{\mathrm{cor}}$ represents a correction factor to translate $\epsilon_{\mathrm{data}}^{\psi (2S)}$ 
to the efficiency for inclusive decays in which the $J/\psi$ is produced at rest. To obtain 
$N_{\mathrm{sel}}$, at least two charged tracks are required for each event. Additionally, the 
momenta of these tracks and the visible energy of each event are restricted in order to eliminate 
Bhabha and di-muon events as well as beam gas interactions and virtual photon-photon collisions. 
The total number of $J/\psi$ decays in the data sample according to Eq.~\eqref{eq:njpsi} is 
determined to be (1.311 $\pm$ 0.011) $\times10^{9}$ events, which results in an uncertainty of 
0.8\%~\cite{YZ14,NJpsi}.

\subsubsection{Background Size}

According to the inclusive MC sample, the total number of background events that contaminate the 
signal is about 1.5\%. These do not include the misreconstructed backgrounds nor the backgrounds 
from $J/\psi$ decays to $\gamma\eta(')$, both of which are addressed in a separate systematic 
uncertainty. Additionally, backgrounds from non-$J/\psi$ decays yield a contamination of 
approximately 0.8\%. Conservative systematic uncertainties equal to 100\% of the background 
contamination are attributed to each of the inclusive MC and continuum background types.

\subsubsection{Uncertainty in the acceptance corrected signal yield}

One of the largest remaining backgrounds after signal isolation and background subtraction is the 
signal mimicking decay of $J/\psi$ to $\omega\pi^{0}$, where the $\omega$ decays to 
$\gamma\pi^{0}$. The nominal method to address this background is to restrict the $\gamma\pi^{0}$ 
invariant mass to exclude the region within 50~MeV/$c^{2}$ of the $\omega$ mass. An alternative 
method is to include an amplitude for the $\omega\pi^{0}$ final state in the analysis. The results of 
this alternative method are quantitatively no different than the nominal results, suggesting that the 
exclusion method is an effective means of addressing the background from $J/\psi$ decays to 
$\omega\pi^{0}$. The difference in the branching fraction using the signal yield for the alternative 
method compared to the nominal method is about 0.8\%.

As discussed above, backgrounds due to $J/\psi$ decays to $\gamma\eta(')$ are addressed in the 
fitting procedure itself by adding an exclusive MC sample to the data, but with a negative weight. The 
systematic uncertainty do to this background is determined by using the data alone. In this way, 
contributions from these backgrounds are treated as signal and inflate the signal yield and 
background size in Eq.~\eqref{eq:br}. The difference in the branching fraction is 0.03\%, which is 
considered a negligible contribution to the systematic uncertainty.

Differences in the effect of the 6C kinematic fit on the data and MC samples may cause a systematic 
difference in the acceptance corrected signal yield. This effect was investigated by loosening the 
restriction on the $\chi^{2}$ from the 6C kinematic fit. For events with a $M_{\pi^{0}\pi^{0}}$ above KK 
threshold, this restriction was relaxed from less than 60 to be less than 125. Events with an invariant 
mass below KK threshold are required to have a $\chi^{2}$ less than 60 rather than less than 20. The 
difference in the branching fraction for the results with the loosened $\chi^{2}$ cut relative to that of 
the nominal results is about 0.1\%.

Another source of systematic uncertainty in the branching fraction is the difference between the 
nominal results and those obtained by applying a model that describes the $\pi\pi$ dynamics. To test 
this effect, a mass dependent fit using interfering Breit-Wigner line shapes was performed. The 
difference in the branching fraction using the acceptance corrected yield of the mass dependent 
analysis compared to the nominal results is about 0.3\%.

The effect of the remaining misreconstructed backgrounds on the results is studied by performing a 
closure test, in which the mass independent amplitude analysis is performed on an exclusive MC 
sample. This MC sample was generated according to the results of a mass dependent amplitude 
analysis of the data and includes the proper angular distributions. After applying the same selection 
criteria that are applied to the data, the MC sample is passed through the mass independent 
analysis. This process is repeated after removing the remaining misreconstructed backgrounds from 
the sample. The difference in the branching fraction between these two methods is 0.01\%. The effect 
of these backgrounds is therefore taken to be negligible.

\begin{table}[ht]
\caption{\label{tab:eff} This table summarizes the systematic uncertainties (in \%) for the branching fraction of radiative $J/\psi$ decays to $\pi^{0}\pi^{0}$.}
\begin{tabular}{c c}
\hline\hline
Source & $J/\psi\rightarrow\gamma\pi^{0}\pi^{0}$ (\%) \\
\hline
Photon detection efficiency & 5.0 \\
Number of $J/\psi$ & 0.8 \\
Inclusive MC backgrounds & 1.5 \\
Non-$J/\psi$ backgrounds & 0.8 \\
$\omega\pi^{0}$ background & 0.8 \\
Kinematic fit $\chi^{2}_{6C}$ & 0.1 \\
Model dependent comparison & 0.3 \\
\hline
Total & 5.4 \\
\hline\hline
\end{tabular}
\end{table}

\subsection{$4^{++}$ amplitude} \label{sec:4pp}

As discussed above, the only $\pi^{0}\pi^{0}$ amplitudes that are accessible in radiative $J/\psi$ 
decays have even angular momentum and positive parity and charge conjugation quantum numbers. 
The mass independent analysis was performed under the assumption that only the $0^{++}$ and 
$2^{++}$ amplitudes are significant. To test this assumption, the analysis was repeated with the 
addition of a $4^{++}$ amplitude. No significant contribution from a $4^{++}$ amplitude is apparent. 

To test the effect of a $4^{++}$ amplitude that may exist in the data and is ignored in the fit, an 
exclusive MC sample was generated using a model constructed from a sum of resonances each 
parameterized by a Breit-Wigner function in a way that optimally reproduces the data.  One of the 
resonances was an $f_{4}(2050)$, which was generated in each component of the $4^{++}$ 
amplitude. The relative size of the $4^{++}$ amplitude was determined from a mass dependent fit to 
the data, in which the $4^{++}$ amplitude contributed 0.43\% to the overall intensity. A mass 
independent amplitude analysis, which did not include a $4^{++}$ amplitude, was then performed on 
this sample. The results indicate that the intensities and phases for the $0^{++}$ and $2^{++}$ 
amplitudes deviate from the input parameters at the order of the statistical errors from the data 
sample in the region between 1.5 and 3.0 GeV/$c^{2}$. Therefore, the systematic error due to the 
effect of ignoring a possible $4^{++}$ amplitude is estimated to be of the same order as the statistical 
errors in the region from 1.5 to 3.0 GeV/$c^{2}$.

\section{Conclusions} \label{Conclusions}

A mass independent amplitude analysis of the $\pi^{0}\pi^{0}$ system in radiative $J/\psi$ decays is 
presented. This analysis uses the world's largest data sample of its type, collected with the BESIII 
detector, to extract a piecewise function that describes the scalar and tensor $\pi\pi$ amplitudes in 
this decay.  While the analysis strategy employed to obtain results has complications, namely 
ambiguous solutions, a large number of parameters, and potential bias in subsequent analyses from 
non-Gaussian effects (see Appendix~\ref{supplement}), it minimizes systematic bias arising from 
assumptions about $\pi\pi$ dynamics, and, consequently, permits the development of dynamical 
models or parameterizations for the data.

In order to facilitate the development of models, the results of the mass independent analysis are 
presented in two ways. The intensities and phase differences for the amplitudes in the fit are 
presented here as a function of $M_{\pi^{0}\pi^{0}}$. Additionally, the intensities and phases for each 
bin of $M_{\pi^{0}\pi^{0}}$ are given in supplemental materials (see Appendix~\ref{supplement}). 
These results may be combined with those of similar reactions for a more comprehensive study of the 
light scalar meson spectrum.  Finally, the branching fraction of radiative $J/\psi$ decays to 
$\pi^{0}\pi^{0}$ is measured to be $(1.15\pm0.05)\times10^{-3}$, where the error is systematic only 
and the statistical error is negligible. This is the first measurement of this branching fraction.

\begin{acknowledgments}
The BESIII collaboration thanks the staff of BEPCII and the IHEP computing center for their strong 
support. This work is supported in part by National Key Basic Research Program of China under 
Contract No. 2015CB856700; National Natural Science Foundation of China (NSFC) under 
Contracts Nos. 11125525, 11235011, 11322544, 11335008, 11425524; the Chinese Academy of 
Sciences (CAS) Large-Scale Scientific Facility Program; the CAS Center for Excellence in Particle 
Physics (CCEPP); the Collaborative Innovation Center for Particles and Interactions (CICPI); Joint 
Large-Scale Scientific Facility Funds of the NSFC and CAS under Contracts Nos. 11179007, 
U1232201, U1332201; CAS under Contracts Nos. KJCX2-YW-N29, KJCX2-YW-N45; 100 Talents 
Program of CAS; INPAC and Shanghai Key Laboratory for Particle Physics and Cosmology; German 
Research Foundation DFG under Contract No. Collaborative Research Center CRC-1044; Istituto 
Nazionale di Fisica Nucleare, Italy; Ministry of Development of Turkey under Contract No. 
DPT2006K-120470; Russian Foundation for Basic Research under Contract No. 14-07-91152; U. S. 
Department of Energy under Contracts Nos. DE-FG02-04ER41291, DE-FG02-05ER41374, DE-
FG02-94ER40823, DESC0010118; U.S. National Science Foundation; University of Groningen 
(RuG) and the Helmholtzzentrum fuer Schwerionenforschung GmbH (GSI), Darmstadt; WCU 
Program of National Research Foundation of Korea under Contract No. R32-2008-000-10155-0; U.S. 
Department of Energy under Grant No. DE-FG02-87ER40365. This research was supported in part 
by Lilly Endowment, Inc., through its support for the Indiana University Pervasive Technology 
Institute, and in part by the Indiana METACyt Initiative. The Indiana METACyt Initiative at IU is also 
supported in part by Lilly Endowment, Inc.
\end{acknowledgments}

\appendix

\section{Amplitudes} \label{sec:amps}

The amplitude for radiative $J/\psi$ decays to $\pi^{0}\pi^{0}$ can be determined in different bases 
depending on the information of interest. For example, in the helicity basis, the amplitude depends on 
the angular momentum and helicity of the $\pi^{0}\pi^{0}$ resonance as well as the angular 
momentum and polarization of the $J/\psi$. It is also possible to relate the amplitudes to radiative 
multipole transitions. Such a basis is useful because it may allow implementation or testing of 
dynamical assumptions. For example, a model may suggest that the E1 radiative transition should 
dominate over the M2 transition.

In the radiative multipole basis, the amplitude for radiative $J/\psi$ decays to $\pi^{0}\pi^{0}$ is given 
by
\begin{widetext}
\begin{equation} \label{eq:A1}
\begin{split}
  U^{M,\lambda_{\gamma}}(\vec{x},s)=&\sum_{j,J_{\gamma},\mu}N_{J_{\gamma}}N_{j}
D_{M,\mu-\lambda_{\gamma}}^{J}(\pi+\phi_{\gamma},\pi-\theta_{\gamma},0)
D_{\mu,0}^{j}(\phi_{\pi},\theta_{\pi},0)
\frac{1}{2}\frac{1+(-1)^{j}}{2} \\
  &\left<J_{\gamma}-\lambda_{\gamma};j\mu|J\mu-\lambda_{\gamma}\right>\frac{1}{\sqrt{2}}[\delta_{\lambda_{\gamma},1}+\delta_{\lambda_{\gamma},-1}P(-1)^{J_{\gamma}-1}]V_{j,J_{\gamma}}(s)
\end{split}
\end{equation}
\end{widetext}
where the parity, total angular momentum, and helicity of the pair of pseudoscalars are given by $P$, 
$j$, and $\mu$, respectively. The $D$ functions are the familiar Wigner D-matrix elements. The 
angular momentum of the photon, $J_{\gamma}$, is related to the nuclear radiative (E1, M2, E3, etc.) 
transitions. Each amplitude is characterized by the angular momentum of the photon and the angular 
momentum of the pseudoscalar pair. The possible values of $J_{\gamma}$ are limited by the 
conservation of angular momentum. The helicity of the radiative photon is given by 
$\lambda_{\gamma}$. The total angular momentum and polarization of the $J/\psi$ are given by $J$ 
and M, respectively. Finally, $N_{j}=\sqrt{\frac{2j+1}{4\pi}}$ is a normalization factor.

The angles ($\phi_{\gamma}$, $\theta_{\gamma}$) are the azimuthal and polar angles of the photon 
in the rest frame of the $J/\psi$, where the direction of the $J/\psi$ momentum defines the x-axis. The 
angles ($\phi_{\pi}$, $\theta_{\pi}$) are the azimuthal and polar angles of one $\pi^{0}$ in the rest 
frame of the $\pi^{0}\pi^{0}$ pair, with the -z axis along the direction of the photon momentum and the 
x-axis is defined by the direction perpendicular to the plane shared by the beam and the z-axis.

Parity is a conserved quantity for strong and electromagnetic interactions. Hence, for $J/\psi$ 
radiative decays, $P=(-1)^{j}$ must be positive. This means that the only intermediate states 
available have $j^{P}=0^{+},2^{+},4^{+}$, etc. Additionally, isospin conservation in strong interactions 
requires $I^{G}$ for the intermediate state to be $0^{+}$ (isoscalar). The complex function 
$V_{j,J_{\gamma}}(s)$ describes the $\pi^{0}\pi^{0}$ production and decay dynamics. In order to 
minimize the model dependence of the mass independent analysis, the dynamical amplitude is 
replaced by a (complex) free parameter in the unbinned extended maximum likelihood fit. Thus, the 
amplitude, in a region around $s$ is given by 
\begin{equation}
  U^{M,\lambda_{\gamma}}(\vec{x},s)=\sum_{j,J_{\gamma}}V_{j,J_{\gamma}} A_{j,J_{\gamma}}^{M,\lambda_{\gamma}}(\vec{x}),
\end{equation}
where
\begin{equation} \label{eq:decamp}
\begin{split}
  A_{j,J_{\gamma}}^{M,\lambda_{\gamma}}(\vec{x})=&
N_{J_{\gamma}}N_{j}
D_{M,\mu-\lambda_{\gamma}}^{J}(\pi+\phi_{\gamma},\pi-\theta_{\gamma},0) \\
&D_{\mu,0}^{j}(\phi_{\pi},\theta_{\pi},0)\frac{1}{2}\frac{1+(-1)^{j}}{2} \\
&\left<J_{\gamma}-\lambda_{\gamma};j\mu|J\mu-\lambda_{\gamma}\right> \\
&\frac{1}{\sqrt{2}}[\delta_{\lambda_{\gamma},1}+\delta_{\lambda_{\gamma},-1}P(-1)^{J_{\gamma}-1}],
\end{split}
\end{equation}
and $\{j,J_{\gamma}\}$ represents the unique amplitudes accessible for the given set of observables, 
$\{M,\lambda_{\gamma}\}$.

\section{Ambiguities} \label{sec:amb}

One of the challenges of amplitude analysis is the issue of ambiguous solutions, two solutions that 
give the same distribution (eg. Ref.~\cite{G01}). In this section, the ambiguous solutions for radiative
$J/\psi$ decays to $\pi^{0}\pi^{0}$ are studied. 

To determine the angular dependence of the amplitudes, it is necessary to write the decay amplitude 
$A_{j,J_{\gamma}}^{M,\lambda_{\gamma}}(\vec{x})$, which is given in Eq.~\eqref{eq:A1}, explicitly as 
a function of the angles ($\phi_{\gamma}$, $\theta_{\gamma}$) and ($\phi_{\pi}$, $\theta_{\pi}$). The 
Clebsch Gordan factors in the amplitude restrict the signs of $\mu$ to be the same as that of 
$\lambda_{\gamma}$. Thus, for $j=2$ and $\lambda_{\gamma}=1$, only the values $\mu=0,1,2$ give 
non-zero amplitude contributions. It is also important to note that the Clebsch Gordan coefficients will 
change sign under $\lambda_{\gamma}\rightarrow-\lambda_{\gamma}$, but only for 
$J_{\gamma}=2$. This will cancel the delta functions in the decay amplitude with the result
\begin{widetext}
\begin{equation} \label{eq:beforeconst}
A_{j,J_{\gamma}}^{M,\lambda_{\gamma}}(\vec{x})= \sum_{\mu}c_{j,\mu}^{J_{\gamma},\lambda_{\gamma}}N_{J_{\gamma}}N_{j}
e^{- i  M(\pi+\phi_{\gamma})} d_{M,\mu-\lambda_{\gamma}}^{1}(\pi-\theta_{\gamma}) \times
e^{- i \mu\phi_{\pi}} d_{\mu,0}^{j}(\theta_{\pi})
\frac{1}{\sqrt{2}}[\delta_{\lambda_{\gamma},1}+\delta_{\lambda_{\gamma},-1}(-1)^{J_{\gamma}-1}]
\end{equation}
\end{widetext}
where the constants $c_{j,\mu}^{J_{\gamma},\lambda_{\gamma}}$ contain the Clebsch-Gordan
coefficients.

Recall that, for the Wigner small $d$-matrix elements, 
\mbox{$d_{1,\pm1}^{1}(\pi-\theta)=d_{1,\mp1}^{1}(\theta)$} and 
\mbox{$d_{1,0}^{1}(\pi-\theta)=d_{1,0}^{1}(\theta)$}. Then, 
\mbox{$d_{M,\mu-\lambda_{\gamma}}^{1}(\pi-\theta)=d_{M,\lambda_{\gamma}-\mu}^{1}(\theta)$}. 
Also, note that the restrictions on $\mu$ mean that the quantity 
\mbox{$\mu-\lambda_{\gamma}=\pm1,0$}. It is also useful to note that 
\mbox{$\mu-\lambda_{\gamma}=\lambda_{\gamma},0,-\lambda_{\gamma}$}, for 
\mbox{$\mu=\pm2,\pm1,0$} respectively. The usefulness of these features appears when one writes 
out the intensity for a given choice of $M$ and $\lambda_{\gamma}$. It is also useful to plug in the 
values for the constants, which are given in Table~\ref{tab:cgcoef}. The intensity in bin $\alpha$ for a 
given choice of observables is then given by
\begin{equation} \label{eq:ambinten}
\begin{split}
  I_{\alpha}(\vec{x})=&\sum_{M,\lambda_{\gamma}}|h_{0}(\theta_{\pi})d_{M,\lambda_{\gamma}}^{1}(\theta_{\gamma})e^{ i \lambda_{\gamma}\phi_{\pi}}+h_{1}(\theta_{\pi})d_{M,0}^{1}(\theta_{\gamma}) \\
  &+h_{2}(\theta_{\pi})d_{M,-\lambda_{\gamma}}^{1}(\theta_{\gamma})e^{- i \lambda_{\gamma}\phi_{\pi}}|^{2}.
\end{split}
\end{equation}
where terms with the same angular dependencies have been grouped according to 
\begin{equation}
\begin{split}
  h_{0}(\theta_{\pi})&=\sqrt{3}V_{0,1}+\sqrt{\frac{3}{2}}(V_{2,1}+\sqrt{5}V_{2,2}+2V_{2,3})d_{0,0}^{2}(\theta_{\pi}) \\
  h_{1}(\theta_{\pi})&=\frac{1}{\sqrt{2}}(3V_{2,1}+\sqrt{5}V_{2,2}-4V_{2,3})d_{1,0}^{2}(\theta_{\pi}) \\
  h_{2}(\theta_{\pi})&=(3V_{2,1}-\sqrt{5}V_{2,2}+V_{2,3})d_{2,0}^{2}(\theta_{\pi})
\end{split}
\end{equation}
and the subscripts on the production amplitudes represent the possible combinations of $j$ and 
$J_{\gamma}$. The following calculations apply for each bin individually.

\begin{table}[ht]
\caption{\label{tab:cgcoef}The constant factors in Eq.~\eqref{eq:beforeconst} are given here.}
\centering
\begin{tabular}{c c c}
  &$c_{0,0}^{J_{\gamma},\lambda_{\gamma}} = 1$ &\\
  $c_{2,0}^{1,\pm1} = \sqrt{\frac{1}{10}}$ & $c_{2,0}^{2,\pm1} = \pm\sqrt{\frac{3}{10}}$ & $c_{2,0}^{3,\pm1} = \sqrt{\frac{6}{35}}$ \\
  $c_{2,1}^{1,\pm1} = \sqrt{\frac{3}{10}}$ & $c_{2,1}^{2,\pm1} = \pm\sqrt{\frac{1}{10}}$ & $c_{2,1}^{3,\pm1} = -\sqrt{\frac{8}{35}}$ \\
  $c_{2,2}^{1,\pm1} = \sqrt{\frac{3}{5}}$ & $c_{2,2}^{2,\pm1} = \mp\sqrt{\frac{1}{5}}$ & $c_{2,2}^{3,\pm1} = \sqrt{\frac{1}{35}}$ \\
\end{tabular}
\end{table}

The amplitudes for which $M$ and $\lambda_{\gamma}$ have the same (opposite) sign, 
$M=\lambda_{\gamma}=\pm1$ ($M=-\lambda_{\gamma}=\pm1$) are related to each other by a sign 
change in the exponential factor. Note that the terms with a factor of $d_{M,0}^{1}$ will change sign 
under $M\rightarrow-M$ and terms with a factor of  $d_{\mu,0}^{j}$ will change sign under 
$\lambda_{\gamma}\rightarrow-\lambda_{\gamma}$. Then, the intensity becomes
\begin{equation}
\begin{split}
  I(\vec{x})=\sum_{M=\lambda_{\gamma}=\pm1}|&h_{0}(\theta_{\pi})d_{1,1}^{1}(\theta_{\gamma})e^{\pm i \phi_{\pi}}+h_{1}(\theta_{\pi})d_{1,0}^{1}(\theta_{\gamma}) \\
  &+h_{2}(\theta_{\pi})d_{1,-1}^{1}(\theta_{\gamma})e^{\mp i \phi_{\pi}}|^{2} \\
  +\sum_{M=-\lambda_{\gamma}=\pm1}|&h_{0}(\theta_{\pi})d_{1,-1}^{1}(\theta_{\gamma})e^{\pm i \phi_{\pi}}-h_{1}(\theta_{\pi})d_{1,0}^{1}(\theta_{\gamma}) \\
  &+h_{2}(\theta_{\pi})d_{1,1}^{1}(\theta_{\gamma})e^{\mp i \phi_{\pi}}|^{2}.
\end{split}
\end{equation}
Note that the term with $h_{1}(\theta_{\pi})$ has changed sign in the opposite combination. The 
properties of small $d$ functions, 
\mbox{$d_{m',m}^{j}(\theta)=(-1)^{m-m'}d_{m,m'}^{j}(\theta)=d_{-m,-m'}^{j}(\theta)$}, have been used 
to write the incoherent pieces of the intensity in the same way.

It is instructive to write the intensity function as
\begin{equation} \label{inten}
\begin{split}
  I(\vec{x})=&f_{0}+f_{1}\cos{2\theta_{\gamma}}+\frac{1}{2}f_{2}\cos{2\phi_{\pi}} \\
  &+\frac{1}{2}f_{3}\sin{2\theta_{\gamma}}\cos{\phi_{\pi}}-\frac{1}{2}f_{4}\cos{2\theta_{\gamma}}\cos{2\phi_{\pi}},
\end{split}
\end{equation}
where
\begin{equation}
\begin{split}
  f_{0}&=\frac{3}{2}[(h_{0})^{2}+(h_{2})^{2}]+(h_{1}^{2}) \\
  f_{1}&=\frac{1}{2}[(h_{0})^{2}+(h_{2})^{2}]-(h_{1})^{2} \\
  f_{2}&=f_{4}=(h_{0}h_{2}^{*}+h_{0}^{*}h_{2}) \\
  f_{3}&=\sqrt{2}(-h_{0}h_{1}^{*}-h_{0}^{*}h_{1}+h_{2}h_{1}^{*}+h_{2}^{*}h_{1}).
\end{split}
\end{equation}

Now, if a set of amplitude couplings, $V$, have been determined by fitting the intensity function in 
Eq.~\eqref{inten} to the data, ambiguities would arise if an alternative set of couplings, $V'$, would 
give the same angular dependence as the original set. In other words, the new set of amplitudes 
must give the same values for the $f_{i}$ functions ($f_{i}'=f_{i}$).

Consider $f_{2}$, which can be written as a linear combination of two quadratic forms
\begin{equation}
  f_{2}=\frac{1}{2}(|h_{0}+h_{2}|^{2}-|h_{0}-h_{2}|^{2}).
\end{equation}
These quadratic forms are given by
\begin{equation} \label{quad}
\begin{split}
  |h_{0}\pm h_{2}|^{2}&=[\cos^{2}{\theta_{\pi}}(3a_{1}\mp a_{3})+(b-a_{1}\pm a_{3})] \\
  &\times[\cos^{2}{\theta_{\pi}}(3a_{1}^{*}\mp a_{3}^{*})+(b^{*}-a_{1}^{*}\pm a_{3}^{*})],
\end{split}
\end{equation}
where for simplicity the production coefficients have been combined into new variables given by
\begin{equation}
\begin{split}
  b&=\sqrt{3}V_{0,1} \\
  a_{1}&=\frac{\sqrt{6}}{4}(V_{2,1}+\sqrt{5}V_{2,2}+2V_{2,3}) \\
  a_{2}&=-\frac{\sqrt{3}}{4}(3V_{2,1}+\sqrt{5}V_{2,2}-4V_{2,3}) \\
  a_{3}&=\frac{\sqrt{6}}{4}(3V_{2,1}-\sqrt{5}V_{2,2}+V_{2,3}).
\end{split}
\end{equation}
Since only the absolute square of each combination of $h_{0}$ and $h_{2}$ appears in the intensity, 
nontrivial ambiguous solutions only appear when the production coefficients are replaced by their 
complex conjugate for one choice of sign in Eq.~\eqref{quad}. That is, if 
$u_{1}=(b,a_{1},a_{2},a_{3})$ and $u_{2}=(b',a_{1}',a_{2}',a_{3}')$, the solutions 
$\{u_{1},u_{2}\}$ and $\{u_{1},u_{2}^{*}\}$ should give consistent values for $h_{0}\pm h_{2}$. This 
requires that either
\begin{equation}
\begin{split}
  h_{0}'+h_{2}'&=h_{0}^{*}+h_{2}^{*} \\
  h_{0}'-h_{2}'&=h_{0}-h_{2}
\end{split}
\end{equation}
or
\begin{equation}
\begin{split}
  h_{0}'+h_{2}'&=h_{0}+h_{2} \\
  h_{0}'-h_{2}'&=h_{0}^{*}-h_{2}^{*}
\end{split}
\end{equation}
Therefore, either
\begin{equation} \label{first}
\begin{split}
  3a_{1}'-a_{3}'&=3a_{1}^{*}-a_{3}^{*} \\
  b'-a_{1}'+a_{3}'&=b^{*}-a_{1}^{*}+a_{3}^{*} \\
  3a_{1}'+a_{3}'&=3a_{1}+a_{3} \\
  b'-a_{1}'-a_{3}'&=b-a_{1}-a_{3}
\end{split}
\end{equation}
or
\begin{equation} \label{second}
\begin{split}
  3a_{1}'-a_{3}'&=3a_{1}-a_{3} \\
  b'-a_{1}'+a_{3}'&=b-a_{1}+a_{3} \\
  3a_{1}'+a_{3}'&=3a_{1}^{*}+a_{3}^{*} \\
  b'-a_{1}'-a_{3}'&=b^{*}-a_{1}^{*}-a_{3}^{*}.
\end{split}
\end{equation}
Both Eq.~\eqref{first} and Eq.~\eqref{second} require that
\begin{equation} \label{eq:constraint}
  \text{Im }{b}=-2\text{ Im }{a_{1}}.
\end{equation}
The difference between Eq.~\eqref{first} and Eq.~\eqref{second} is a sign change for imaginary part 
of each amplitude. This difference is equivalent to the trivial ambiguities discussed in section 
\ref{sec:ambiguities}. Let us choose the phase convention given by Eq.~\eqref{first}. Finally, 
invariance of $f_{1}$, given the conditions above, requires that $a_{2}'=a_{2}$. 

Using the conditions in Eq.~\eqref{first} and the constraint $a_{2}'=a_{2}$, the alternate set of 
solutions can be written in terms of the original set as
\begin{equation} \label{eq:pred}
\begin{split}
  \text{Re }{V_{0,1}'}&=\text{Re }{V_{0,1}} \\
  \text{Im }{V_{0,1}'}&=-\frac{1}{3\sqrt{2}}(3\text{ Im }{V_{2,1}}-\sqrt{5}\text{ Im }{V_{2,2}}+\text{ Im }{V_{2,3}}) \\
  \text{Re }{V_{2,1}'}&=\text{Re }{V_{2,1}} \\
  \text{Im }{V_{2,1}'}&=\text{Im }{V_{2,1}}+\frac{2\sqrt{5}}{3}\text{ Im }{V_{2,2}}+\frac{5}{6}\text{ Im }{V_{2,3}} \\
  \text{Re }{V_{2,2}'}&=\text{Re }{V_{2,2}} \\
  \text{Im }{V_{2,2}'}&=-\text{Im }{V_{2,2}}-\frac{\sqrt{5}}{2}\text{ Im }{V_{2,3}} \\
  \text{Re }{V_{2,3}'}&=\text{Re }{V_{2,3}} \\
  \text{Im }{V_{2,3}'}&=\text{Im }{V_{2,3}}.
\end{split}
\end{equation}
Note that the last two lines of Eq.~\eqref{eq:pred} indicate that the ambiguous solution for the
$2^{++}$ E3 amplitude is redundant with the original solution. That is, the $2^{++}$ E3 amplitude 
does not exhibit multiple solutions.

In a practical sense, these results are useful to compare the mathematical predictions to what is 
found experimentally. Essentially, the predicted ambiguous partner for a set of fit results in a given 
bin may be calculated in the following way. First, the results must be rotated in phase space such that 
the condition in Eq.~\eqref{eq:constraint} is satisfied. Next, the ambiguous partner may be 
determined using Eq.~\eqref{eq:pred}. Finally, this predicted solution must be rotated back into the 
original phase convention. Now, the predicted ambiguous partner may be compared with the 
experimentally determined fit results. Studies show that the mathematically predicted ambiguities 
match those found experimentally.

\section{Supplemental Materials} \label{supplement}

In addition to the figures presented here, the results of the mass independent analysis in each bin of 
$M_{\pi^{0}\pi^{0}}$ are included in the supplemental materials~\cite{SUPP}. This includes the 
intensities of each amplitude and the three phase differences for each bin of $M_{\pi^{0}\pi^{0}}$. The 
two ambiguous solutions of the nominal results are separated into two text files, while one additional 
text file contains the alternate results in the region where they are not redundant with the nominal 
results. Note that these results contain only statistical errors.

It is important to reiterate that errors reported in the supplemental results (and in the figures in the 
text) are derived from the covariance matrix of the fit parameters.  That is, they are valid in the 
Gaussian limit, a limit that cannot be guaranteed for all parameters in the analysis.  Therefore the use 
of these results in a subsequent fit to parameters of interest cannot be expected to produce 
statistically rigorous values of the parameters.  Likewise a $\chi^2$ or likelihood-ratio test of a model 
describing the results cannot be rigorously constructed.

An attempt to quantify the potential systematic bias in subsequent analyses was made as follows.  
First, a sample of MC with equivalent statistical precision to the data was generated using a model 
consisting of a coherent sum of Breit-Wigner resonances in a way that best approximates the data.  A 
mass independent amplitude analysis was performed on this MC sample using the same procedure 
that was applied to the actual data reported in this analysis.  The results of this mass independent 
analysis of the MC sample were then fit with a Breit-Wigner model, the same model with which they 
were generated, where the couplings of the Breit-Wigner distributions in the model were allowed to 
float as free parameters.  While most fit parameters exhibited typical Gaussian fluctuations about their 
known input values, there were some non-Gaussian outliers.  About one-third of the parameters 
exhibited deviations from input at or above the three sigma level. In comparison with a mass 
dependent analysis, in which the Breit-Wigner model is directly fit to the same mock data, the 
parameter errors in the model fit to the MI results were generally larger, typically within a factor of two, 
but in some cases by up to a factor of ten.

To probe the scale of the systematic deviations of the fitted values from the true input values used to 
generate our MC sample, for each amplitude we used the true value of
the coupling instead of the fitted value and computed (1) the total intensity integrated over all phase 
space and (2) the fit fraction (ratio of individual amplitude intensity to total intensity).  We observe the 
deviations in (1) to be at or below the 1\% level for all amplitudes and deviations in (2) to be at or 
below 2\% on an absolute scale for all amplitudes.  For small amplitudes, this means that relative 
deviations in intensity may occur at a level of 10-90\%. This suggests validity and precision at a level 
sufficient for model development; however, rigorous values for any model parameters can only be 
reliably obtained by
fitting the given model directly to the data.

\end{document}